\documentclass[aps,pra,twocolumn,showpacs,floatfix]{revtex4}
\usepackage{epsfig}
\usepackage{graphicx}
\usepackage{dcolumn}
\usepackage{amsthm,amsmath}

\begin{document}
\title{ Magic wavelengths for trapping the alkali-metal atoms with circularly polarized light}
\author{B. K. Sahoo }
\affiliation{Theoretical Physics Division, Physical Research Laboratory, Navrangpura, Ahmedabad-380009, India}
\author{Bindiya Arora \footnote{Email: arorabindiya@gmail.com}} 
\affiliation{Department of Physics, Guru Nanak Dev University, Amritsar, Punjab-143005, India} 
\date{Received date; Accepted date}
 
\begin{abstract}
Magic wavelengths for the Li, Na and K alkali atoms are determined using the circulalrly
polarized light for the $ns-np_{1/2,3/2}$ transitions, with $n$ denoting their ground state
principal quantum numbers, by studying their differential ac dynamic polarizabilities. 
These wavelengths for all possible sub-levels are given  linearly as well as
circulalrly polarized lights and are further compared with the available results for the linearly 
polarized light. The present study suggests that it is possible to carry out state 
insensitive trapping of different alkali atoms using the circularly polarized light.
\end{abstract}

\maketitle

\section{Introduction}\label{sec1}
It has been known for some time that atoms can be trapped and manipulated by the gradient forces of light waves \cite{1,2,3}.
However, for any two internal states of an atom, the Stark shifts caused due to the trap
light are different which affects the fidelity of the experiments \cite{QC2,katori1}.
Katori \textit{et al.} \cite{katori2} suggested a solution to this problem that the trapping laser can be tuned to
wavelength, ``$\lambda_{\rm{magic}}$", where the differential ac Stark shifts of the transition 
vanishes.
Knowledge of magic wavelengths is necessary in many areas of physics. 
In particular, these wavelengths have unavoidable application for atomic clocks and quantum 
computing~\cite{AC1,QC1}. 
For example, a major concern for the accuracy of optical lattice clocks is the ability to cancel the large light shifts created by the trapping potential of the lattice.
Similarly, for most of the quantum computational schemes, it is often desirable to optically trap the neutral atoms
without affecting the internal energy-level spacing for the atoms.

Over the years, there has been quite a large number of calculations of the magic wavelengths of alkali-metal atoms for linearly polarized traps \cite{ludlow,kimble1,arora1}, with
rather fewer calculations of the magic wavelength for the circularly polarized 
traps. Moreover, as stated in \cite{arora1}, the linearly polarized lattice scheme
offers only a few cases in which the magic wavelengths are of experimental relevance. Therefore, we would like to explore
the idea of using the circularly polarized light. Using the circularly polarized light may be advantageous owing to the 
dominant role played by vector polarizabilities in estimating the ac Stark shifts \cite{sahoo-arora2,derevianko1,zeeman1}.
These polarizability contributions are absent in the linearly polarized
light. Recently, we had investigated the magic wavelengths for the circularly polarized 
light in rubidium (Rb) atom and found spectacular results that can lead to state
insensitive trapping of Rb atoms using this light \cite{sahoo-arora2}. In this
paper, we aim at searching for magic wavelengths in the Li, Na, and K atoms due to circularly polarized light
and also compare our results for linearly polarized light with other available results.

\section{Theory}\label{sec2}
The energy shift of any state $v$ of an atom placed in a frequency-dependent ac electric field $\mathcal{E}(\omega)$, can be 
estimated from the time-independent perturbation theory at the second order perturbation level as
	\begin{equation}
		\Delta E_v \simeq -\frac{1}{2}\alpha_v(\omega) \mathcal{E}^2, \label{eq-1}
	\end{equation}	
where $\alpha_{v}(\omega)$ is the dynamic dipole polarizability of the atom in the $v^{th}$ state and can be expressed as
	\begin{equation}
\alpha_v (\omega) = - \sum_{k \neq v} ({p^*})_{vk}({p})_{kv} \left [ \frac {1}{\delta E_{vk} + \omega } + \frac {1}{\delta E_{vk} - \omega } \right ],
	\end{equation}
where $({p})_{kv}={\left\langle \psi_k\left|D\right|\psi_v\right\rangle}$ is the electric-dipole matrix element.
In a more conventional form, the dipole polarizability $\alpha(\omega)$ can be decomposed into three components as
\begin{eqnarray}
\alpha_v (\omega) &=& \alpha_v^0(\omega) + \mathcal{A} \cos{\theta_k} \frac{m_j}{j} \alpha_v^1(\omega) \nonumber \\ && + \left \{ \frac{3 \cos^2{\theta_p} -1}{2} \right \} \frac{3m_j^2 - j(j+1)}{j(2j-1)} \alpha_v^2(\omega),
\label{cpl}
\end{eqnarray}
which separates out the $m_j$ dependent and independent components. The $m_j$ independent parameters 
$\alpha_0(\omega)$, $\alpha_1(\omega)$ and $\alpha_2(\omega)$ are known as scalar, vector and tensor 
polarizabilities, respectively. They are given in terms of the reduced matrix dipole matrix elements 
as \cite{manakov}
\begin{eqnarray}
\alpha_v^{0}(\omega) &=& \frac{1}{{3(2j_v+1)} }\sum_{j_k}
|\langle \psi_v \parallel D\parallel \psi_k \rangle|^{2} \nonumber \\ && \times \left [ \frac{1}{\delta E_{kv}+\omega}+\frac{1}{\delta E_{kv}-\omega} \right ],\label{eq-scalar}
\end{eqnarray}
\begin{eqnarray}
\alpha_v^{1}(\omega) &=& - \sqrt { \frac{6j_v} {(j_v+1)(2j_v+1)} }
\sum_{j_k} \left \{ \begin{array}{ccc} j_v & 1 & j_v \\ 1 & j_k & 1 \end{array} \right \} \nonumber \\ && (-1)^{j_v+j_{k}+1} 
 | \langle \psi_v \parallel D \parallel \psi_k \rangle|^{2} \nonumber \\ && \times \left [ \frac{1}{\delta E_{kv}+\omega}-\frac{1}{\delta E_{kv}-\omega} \right ] \label{eq-vector}
\end{eqnarray}
and
\begin{eqnarray}
\alpha_v^{2}(\omega) &=& -2 \sqrt { \frac{5j_v(2j_v-1)} {6(j_v+1)(2j_v+1)(2j_v+3)} } \nonumber \\ && 
\sum_{j_k} \left \{ \begin{array}{ccc} j_v & 2 & j_v \\ 1 & j_k & 1 \end{array} \right \} (-1)^{j_v+j_{k}+1} 
 | \langle \psi_v \parallel D \parallel \psi_k \rangle|^{2}  \nonumber \\ && \times \left [ \frac{1}{\delta E_{kv}+\omega}+\frac{1}{\delta E_{kv}-\omega} \right ]\label{eq-tensor}.
\end{eqnarray}

In the above expression $\mathcal{A}$, $\theta_k$ and $\theta_p$ define degree of circular polarization, 
angle between wave vector of the electric field and $z$-axis and angle between the direction of polarization 
and $z$-axis, respectively. Without the loss of generality, it is assumed that the considered frequencies 
($\omega$s) are several line-widths off from the resonance lines and $\mathcal{A}=1$ for the right-handed
and $\mathcal{A}=-1$ for the left-handed circularly polarized light. In the absence of the
magnetic field (or in weak magnetic field), we approximate $\cos(\theta_k ) = \cos( \theta_p ) = 1$. 

The differential ac Stark shift for a transition is defined as the difference between the Stark shifts
of individual levels which are further calculated from the frequency-dependent polarizabilities:
\begin{eqnarray}
 \delta(\Delta E)_{np_i-ns} &=&  \Delta E_{np_i} - \Delta E_{ns} \nonumber \\
   &=& \frac{1}{2} \mathcal{E}^2 (\alpha_{ns} - \alpha_{np_i}),
\end{eqnarray}
where we have used the total polarizabilities of the respective states and $i=1/2,3/2$. Since
the external electric field $\mathcal{E}$ is arbitrary, we can locate the frequencies or wavelengths where
$\alpha_{np_i}=\alpha_{ns}$, for an atom for the null differential ac Stark shifts which gives the value of magic wavelengths. 
In other words, the crossing between the two polarizabilities at various values 
of wavelengths will correspond to $\lambda_{\rm{magic}}$. As pointed out in the begining, it will be experimentally convenient to trap atoms at these wavelengths.

\section{Procedure for calculations}\label{sec3}
The scalar, vector and tensor polarizabilities can be written using sum-over intermediate
states as
\begin{eqnarray}
\alpha_v^{\lambda} &=& \sum_{j_k} C_{v,k}^{\lambda} \frac{|\langle j_v || D || j_k \rangle |^2} {E_v - E_k},
\end{eqnarray}
where $\lambda=0, \ 1$ and $2$ represents for scalar, vector and tensor polarizabilities
and $C_{v,k}^{\lambda}$ are their corresponding angular coefficients. In order to apply
this formula, it is necessary to determine intermediate $k$ states explicitly. Therefore,
contributions from the intermediate states involving core orbitals cannot be determined
in this procedure. For a practical approach, we divide contributions to $\alpha_v$ as
\begin{eqnarray}
\alpha_v^{\lambda}  &=& \alpha_v^{\lambda}(c) + \alpha_v^{\lambda}(vc) + \alpha_v^{\lambda}(v),
\end{eqnarray}
by expressing wave functions of the state $v$ as a closed core with the corresponding
valence orbital so that $\alpha_v^{\lambda}(c)$ and $\alpha_v^{\lambda}(vc)$ account
contributions from the intermediate states involving core orbitals and $\alpha_v^{\lambda}(v)$
take care of the contributions from the excited states involving the virtual orbitals. As a
result, $\alpha_v^{\lambda}(v)$ contributions will be the dominant ones among them and
and can be estimated determining the important low-lying intermediate states explicitly. 
Contributions from $\alpha_v^{\lambda}(c)$, $\alpha_v^{\lambda}(vc)$ and higher intermediate
states with the virtual orbitals (given as $tail$ contributions) are obtained using the
third order many-body perturbation theory (MBPT(3) method) in the Lewis-Galgarno 
approach \cite{sahoo-arora2, arora-sahoo1}.

To determine contributions to $\alpha_v^{\lambda}(v)$ from the low-lying intermediate
states involving virtual orbitals, we express atomic wave functions of a given state 
with a colsed core and a valence orbital $k$ as
\begin{eqnarray}
| \Phi_k \rangle =a_k^{\dagger}|\Phi_0\rangle,
\end{eqnarray}
where $|\Phi_0\rangle$ is the Dirac-Fock (DF) wave function for the closed core and $a_k^{\dagger}$
correspond to the attachment of the $k$ valence orbital to the core. The wave function of the
exact state is then expressed in the coupled-cluster (CC) theory framework as \cite{rcc}
\begin{eqnarray}
|\Psi_v \rangle &=& e^T \{1+S_v\} |\Phi_v \rangle ,
\label{cc1}
\end{eqnarray}
where $T$ and $S_v$ operators account coorelation effects from core and core with valence
orbitals, respectively. Amplitudes of these operators are obtained using the Dirac-Coulomb
(DC) Hamiltonian and the detailed procedures are explained else where (e.g. refer to \cite{mukherjee,sahoo2}). 

\begin{table}
\caption{\label{e1mat} Absolute values of E1 matrix elements in Li, Na, and K atoms in $ea_0$ using  CCSD(T) methods. Uncertainties in the CCSD(T) results are given in the parentheses.}
\begin{ruledtabular}
\begin{tabular}{lclc}
Transition & CCSD(T) & Transition & CCSD(T)  \\
\hline
Li \\
\hline
$2s_{1/2} \rightarrow 2p_{1/2}$  &  3.318(4)& $2p_{1/2} \rightarrow 5d_{3/2}$  & 1.212(1) \\ 
$2s_{1/2} \rightarrow 3p_{1/2}$  & 0.182(2)&$2p_{1/2} \rightarrow 6d_{3/2}$  & 0.767(1) \\
$2s_{1/2} \rightarrow 4p_{1/2}$  & 0.159(2) & $2p_{1/2} \rightarrow 7d_{3/2}$  & 0.566(5)\\
$2s_{1/2} \rightarrow 5p_{1/2}$  & 0.119(4)&$2p_{3/2} \rightarrow 3s_{1/2}$  & 3.445(3)\\
$2s_{1/2} \rightarrow 6p_{1/2}$  & 0.092(2)& $2p_{3/2} \rightarrow 4s_{1/2}$  & 0.917(2)\\
$2s_{1/2} \rightarrow 7p_{1/2}$  & 0.072(1) & $2p_{3/2} \rightarrow 5s_{1/2}$  & 0.493(2)\\
$2s_{1/2} \rightarrow 2p_{3/2}$  & 4.692(5)& $2p_{3/2} \rightarrow 6s_{1/2}$  & 0.326(2)\\
$2s_{1/2} \rightarrow 3p_{3/2}$  & 0.257(2)&$2p_{3/2} \rightarrow 7s_{1/2}$  & 0.233(2)\\
$2s_{1/2} \rightarrow 4p_{3/2}$  & 0.225(2)&$2p_{3/2} \rightarrow 3d_{3/2}$  & 2.268(2)\\
$2s_{1/2} \rightarrow 5p_{3/2}$  & 0.169(4)&$2p_{3/2} \rightarrow 4d_{3/2}$  & 0.863(1) \\
$2s_{1/2} \rightarrow 6p_{3/2}$  & 0.130(2)& $2p_{3/2} \rightarrow 5d_{3/2}$  & 0.502(1)\\
$2s_{1/2} \rightarrow 7p_{3/2}$  & 0.102(1)& $2p_{3/2} \rightarrow 6d_{3/2}$  & 0.344(1)\\
$2p_{1/2} \rightarrow 3s_{1/2}$  & 2.436(3)& $2p_{3/2} \rightarrow 7d_{3/2}$  & 0.253(4)\\
$2p_{1/2} \rightarrow 4s_{1/2}$  & 0.648(3)& $2p_{3/2} \rightarrow 3d_{5/2}$  & 6.805(1)\\
$2p_{1/2} \rightarrow 5s_{1/2}$  & 0.349(3)& $2p_{3/2} \rightarrow 4d_{5/2}$  & 2.589(1)\\
$2p_{1/2} \rightarrow 6s_{1/2}$  & 0.231(2)& $2p_{3/2} \rightarrow 5d_{5/2}$  & 1.505(1) \\
$2p_{1/2} \rightarrow 7s_{1/2}$  & 0.165(2)& $2p_{3/2} \rightarrow 6d_{5/2}$  & 1.031(1)\\
$2p_{1/2} \rightarrow 3d_{3/2}$  & 5.072(1)& $2p_{3/2} \rightarrow 7d_{5/2}$  & 0.746(5)\\
$2p_{1/2} \rightarrow 4d_{3/2}$  & 1.929(1)& \\
\hline
Na \\
\hline
$3s_{1/2} \rightarrow 3p_{1/2}$   & 3.545(3) & $3p_{1/2} \rightarrow 5d_{3/2}$  & 0.997(2)\\ 
$3s_{1/2} \rightarrow 4p_{1/2}$   & 0.304(2)&$3p_{1/2} \rightarrow 6d_{3/2}$   & 0.645(1)\\
$3s_{1/2} \rightarrow 5p_{1/2}$   & 0.107(1)&$3p_{1/2} \rightarrow 7d_{3/2}$   & 0.460(1) \\
$3s_{1/2} \rightarrow 6p_{1/2}$   & 0.056(2)& $3p_{3/2} \rightarrow 4s_{1/2}$   & 5.070(4) \\
$3s_{1/2} \rightarrow 7p_{1/2}$   & 0.035(2)&$3p_{3/2} \rightarrow 5s_{1/2}$   & 1.072(2)\\
$3s_{1/2} \rightarrow 8p_{1/2}$   & 0.026(2)&$3p_{3/2} \rightarrow 6s_{1/2}$   & 0.553(2)\\
$3s_{1/2} \rightarrow 3p_{3/2}$   & 5.012(4)& $3p_{3/2} \rightarrow 7s_{1/2}$   & 0.360(1)\\
$3s_{1/2} \rightarrow 4p_{3/2}$   & 0.434(2)& $3p_{3/2} \rightarrow 8s_{1/2}$   & 0.255(1)\\
$3s_{1/2} \rightarrow 5p_{3/2}$   & 0.153(2)&$3p_{3/2} \rightarrow 3d_{3/2}$   & 3.048(3)\\
$3s_{1/2} \rightarrow 6p_{3/2}$   & 0.081(2)&$3p_{3/2} \rightarrow 4d_{3/2}$   & 0.856(2)\\
$3s_{1/2} \rightarrow 7p_{3/2}$   & 0.051(2)&$3p_{3/2} \rightarrow 5d_{3/2}$  & 0.445(2)\\
$3s_{1/2} \rightarrow 8p_{3/2}$   & 0.037(2)&$3p_{3/2} \rightarrow 6d_{3/2}$   & 0.288(1)\\
$3p_{1/2} \rightarrow 4s_{1/2}$   & 3.578(4)&$3p_{3/2} \rightarrow 7d_{3/2}$   & 0.205(1)\\
$3p_{1/2} \rightarrow 5s_{1/2}$   & 0.758(3)&$3p_{3/2} \rightarrow 3d_{5/2}$   & 9.144(4) \\
$3p_{1/2} \rightarrow 6s_{1/2}$   & 0.391(2)& $3p_{3/2} \rightarrow 4d_{5/2}$   & 2.570(3)\\
$3p_{1/2} \rightarrow 7s_{1/2}$   & 0.255(2)& $3p_{3/2} \rightarrow 5d_{5/2}$   & 1.336(2)\\
$3p_{1/2} \rightarrow 8s_{1/2}$   & 0.180(1)&$3p_{3/2} \rightarrow 6d_{5/2}$   & 0.864(2) \\
$3p_{1/2} \rightarrow 3d_{3/2}$   & 6.807(3)&$3p_{3/2} \rightarrow 7d_{5/2}$   & 0.606(2)  \\
$3p_{1/2} \rightarrow 4d_{3/2}$   & 1.916(2)& \\
\hline
K \\
\hline
$4s_{1/2} \rightarrow 4p_{1/2}$   & 4.131(20) & $4p_{1/2} \rightarrow 6d_{3/2}$   & 0.293(5)\\ 
$4s_{1/2} \rightarrow 5p_{1/2}$   & 0.282(6)& $4p_{1/2} \rightarrow 7d_{3/2}$   & 0.261(4) \\
$4_{1/2} \rightarrow 6p_{1/2}$    & 0.087(5)&$4p_{1/2} \rightarrow 8d_{3/2}$   & 0.221(4) \\
$4s_{1/2} \rightarrow 7p_{1/2}$   & 0.041(5)&$4p_{3/2} \rightarrow 5s_{1/2}$   & 5.524(10)\\
$4s_{1/2} \rightarrow 8p_{1/2}$   & 0.023(3) &$4p_{3/2} \rightarrow 6s_{1/2}$   & 1.287(10)\\
$4s_{1/2} \rightarrow 9p_{1/2}$   & 0.016(3) &$4p_{3/2} \rightarrow 7s_{1/2}$   & 0.677(6)\\
$4s_{1/2} \rightarrow 4p_{3/2}$   & 5.841(20) &$4p_{3/2} \rightarrow 9s_{1/2}$  & 0.317(5) \\
$4s_{1/2} \rightarrow 5p_{3/2}$   & 0.416(6)&$4p_{3/2} \rightarrow 10s_{1/2}$  & 0.242(5)\\
$4s_{1/2} \rightarrow 6p_{3/2}$   & 0.132(6)&$4p_{3/2} \rightarrow 3d_{3/2}$   & 3.583(20)\\
$4s_{1/2} \rightarrow 7p_{3/2}$   & 0.064(5)&$4p_{3/2} \rightarrow 4d_{3/2}$   &  0.088(5)\\
$4s_{1/2} \rightarrow 8p_{3/2}$   & 0.038(3)&$4p_{3/2} \rightarrow 5d_{3/2}$   &  0.124(5)\\
$4s_{1/2} \rightarrow 9p_{3/2}$   & 0.027(3)&$4p_{3/2} \rightarrow 6d_{3/2}$   & 0.135(5)\\
$4p_{1/2} \rightarrow 5s_{1/2}$   & 3.876(10)&$4p_{3/2} \rightarrow 7d_{3/2}$   & 0.119(4)\\
$4p_{1/2} \rightarrow 6s_{1/2}$   & 0.909(10)&$4p_{3/2} \rightarrow 8d_{3/2}$   & 0.101(3)\\
$4p_{1/2} \rightarrow 7s_{1/2}$   & 0.479(5)&$4p_{3/2} \rightarrow 3d_{5/2}$   & 10.749(50)\\
$4p_{1/2} \rightarrow 8s_{1/2}$   & 0.316(5)&$4p_{3/2} \rightarrow 4d_{5/2}$   & 0.260(5)\\
$4p_{1/2} \rightarrow 9s_{1/2}$   & 0.225(3)&$4p_{3/2} \rightarrow 5d_{5/2}$   & 0.374(5)\\
$4p_{1/2} \rightarrow 10s_{1/2}$  & 0.171(3)& $4p_{3/2} \rightarrow 6d_{5/2}$   & 0.404(5)\\
$4p_{1/2} \rightarrow 3d_{3/2}$   & 7.988(40)& $4p_{3/2} \rightarrow 7d_{5/2}$   & 0.356(5)\\
$4p_{1/2} \rightarrow 4d_{3/2}$   & 0.220(5) & $4p_{3/2} \rightarrow 8d_{5/2}$   & 0.286(5)\\
$4p_{1/2} \rightarrow 5d_{3/2}$   & 0.264(5)&\\
\end{tabular}   
\end{ruledtabular}

\end{table}

\begin{table}
\caption{\label{pol}
Comparision of polarizabilities of the ground and first two excited states in Li, Na and K (a.u.) 
from this work with the results reported in \cite{safronova-li}$^a$, \cite{li-exp}$^b$, 
\cite{pol-andrei}$^c$, \cite{arora1}$^d$, \cite{na-exp}$^e$, \cite{windholzm}$^f$, $^g$Ref.~\cite{k-exp},$^h$derived from 
Ref.~\cite{hunter1} D1 line Stark shift measurements and recommended values for ground state polarizability from 
Ref.~\cite{pol-andrei},$^i$Ref.~\cite{krenn}.}.

\begin{ruledtabular}
\begin{tabular}{lcccc}
   Contributions & \multicolumn{1}{c}{$ns$}  & \multicolumn{1}{c}{$np_{1/2}$} & \multicolumn{2}{c}{$np_{3/2}$} \\
            from  &  $\alpha_v^0$ &  $\alpha_v^0$ &  $\alpha_v^0$ & $\alpha_v^2$ \\
	      \hline
\multicolumn{4}{l}{Li ($n=2$)} \\
\hline
              $2s$     & - & -54.041(4) & -54.031(5) & 54.031(5)\\
              $2p_{1/2}$ &54.041(4) &-&-&- \\
              $2p_{3/2}$ &108.06(1) &-&-&- \\
              $3s$  &- &35.288(2) & 35.288(2) & -35.288(2) \\
              $3p_{1/2}$ &0.078 &-&-&- \\
              $3p_{3/2}$ &0.156 &-&-&- \\
              $3d_{3/2}$ &-&114.901(2)&11.488(1)&9.190 \\
              $3d_{5/2}$ &- &-&103.419(2)&-20.684 \\
              $4s$   &- &1.528()&1.530&-1.530 \\
              $4p_{1/2}$ &0.051&-&-&- \\
              $4p_{3/2}$ &0.102&-&-&- \\
              $4d_{3/2}$ & -&12.534(1)&1.254&-2.258 \\
              $4d_{5/2}$ & -&-&11.289&-2.258 \\
              $\alpha_v(v)$ & 0.154 & 7.63 & 6.931 & -0.627\\
              $\alpha_v(cv)$ &$\sim 0$ & $\sim 0$& $\sim 0$ & $\sim 0$\\
              $\alpha_v(tail)$ & 1.2(6) &10(5)& 10(5)&-1.7(8) \\
Present       & 164.1(6) &  128        & 127  &1.5 \\
Others     & 164.16(5)$^a$ & 126.97(5)$^a$& 126.98(5)$^a$ & 1.610(26)$^a$ \\
Expt.   &   164.2(11)$^b$ & -&- &- \\
\hline
\multicolumn{4}{l}{Na ($n=3$)} \\
\hline
             {\bf 3s}     &- &-53.60(7) &-53.53(7) & 53.53(7) \\
             ${\bf 3p_{1/2}}$ &53.60(7) &- &-&- \\
               ${\bf3p_{3/2}}$ &107.06(15) &-&-&- \\
              $4s$  &- &106.625(9)&107.255(9)&-107.255(9) \\
              $3d_{3/2}$ &- &277.47(2)&27.856(2)&22.285(1) \\
              $3d_{5/2}$ &- &- &250.71(2)&-50.141(4) \\
              $4p_{1/2}$ & 0.223&-&-&- \\
              $4p_{3/2}$ & 0.455&- &-&- \\
              $5s$   & -& 2.588&2.591&-2.591\\
              $5p_{1/2}$ &0.024 &-&-&- \\
              $5p_{3/2}$ &0.049 &-&-&- \\
              $4d_{3/2}$ &- &15.266(1)&1.525&1.220 \\
              $4d_{5/2}$ &- &-&13.747&-2.749 \\
              $\alpha_v(v)$ & 0.030&6.647&6.616&-1.479 \\
              $\alpha_v(cv)$ &$\sim 0$ & $-1.3 $x $10^{-4}$  & $-2.6$ x  $10^{-4}$ & $\sim 0$\\
              $\alpha_v(tail)$ & 0.08(4) & 5(3)& 5(3) &-1.5(7)\\
Present   & 162.4(2)     &    361     &   362      &-88\\
Others & 162.6(3)$^c$   & 359.9$^d$      & 361.6$^d$ & -88.4$^d$ \\
Expt.  &   162.7(8)$^e$& 359.2(6)$^e$ & 360.4(7)$^e$ & -88.3(4)$^f$\\
\hline
\multicolumn{4}{l}{K ($n=4$)} \\
\hline
             ${\bf 4s}$     &- &-94.8(2)&-94.3(3)&94.3(3) \\
               ${\bf 4p_{1/2}}$ &94.8(2) &-&-&- \\
             $ {\bf 4p_{3/2}}$ & 188.7(5)&-&-&- \\
              $5s$  & -&0.027&139.81(3)&-139.81(3) \\
              $3d_{3/2}$ &- &545.9(4)&55.29(2)&44.23(2) \\
              $3d_{5/2}$ & -&-&497.7(5)&-99.54(10) \\
              $5p_{1/2}$ & 0.236&-&-&- \\
              $5p_{3/2}$ & 0.512&-&-&- \\
              $4d_{3/2}$ &- &0.246&0.020&0.016 \\
              $4d_{5/2}$ &- &-&0.172&-0.035 \\
              $6s$   &- &4.179(1)&4.205(1)&-4.205(1) \\
              $6p_{1/2}$ &0.019 &-&-&- \\
              $6p_{3/2}$ & 0.044&-&-&- \\
              $5d_{3/2}$ &- &0.296&0.033&0.026 \\
              $5d_{5/2}$ &- &-&0.299&-0.060 \\
              $\alpha_v(v)$ & 0.020&2.421&2.444&-3.160\\
              $\alpha_v(cv)$ & -0.13 & $\sim 0$& $\sim 0$ & $\sim 0$\\
              $\alpha_v(tail)$ & 0.06(3)&6.1(6)&6.2(6)&-2.1(4)\\
Present   & 289.8(6)      &   605.3    &   616.0      &-107.5\\
Others &  290.2(8)$^c$ & 602$^d$  &  613$^d$&  -109$^d$\\
Expt.  &290.58(1.42)$^g$& 606.7(6)$^h$ & 614(10)$^i$& 107(2)$^i$ \\
\end{tabular}
\end{ruledtabular}
\end{table}

We calculate the reduced matrix elements of $D$ between states $| \Psi_f \rangle$ and 
$| \Psi_i \rangle$ after obtaining from the above procedure that to be used in the 
sum-over-states approach as
expression
\begin{eqnarray}
\langle \Psi_f || D || \Psi_i \rangle &=& \frac{\langle \Phi_f || \{ 1+ S_f^{\dagger}\} \overline{D } \{ 1+ S_i\} ||\Phi_i\rangle}{ \sqrt{{\cal N}_f {\cal N}_i}}, 
\end{eqnarray}
where $\overline{ D}=e^{T^{\dagger}} D e^T$ and ${\cal N}_v = \langle \Phi_v |
e^{T^{\dagger}} e^T + S_v^{\dagger} e^{T^{\dagger}} e^T S_v |\Phi_v\rangle$. Detailed calculation 
procedures of these expressions are discussed elsewhere \cite{mukherjee,sahoo2}.

\begin{figure}[htp]
 \includegraphics[scale=0.6]{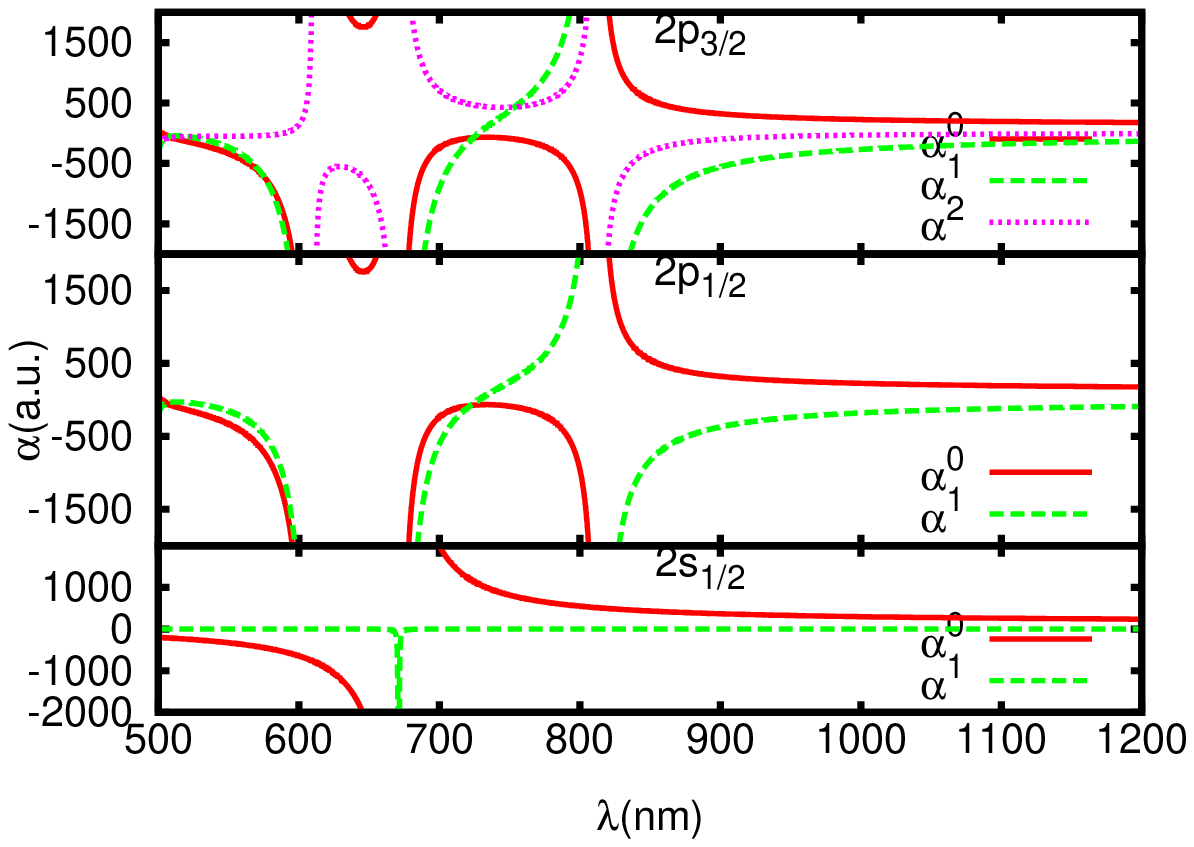}
  \caption{(color online)Scalar $\alpha^0$, vector $\alpha^1$ and tensor $\alpha^2$ polarizability (in a.u.) for $2s_{1/2}$, $2p_{1/2}$ and $2p_{3/2}$ state of Li atom as a function of wavelength (in n.m.)}
  \label{figli}     
\end{figure}

\begin{figure}[htp]
  \includegraphics[scale=0.6]{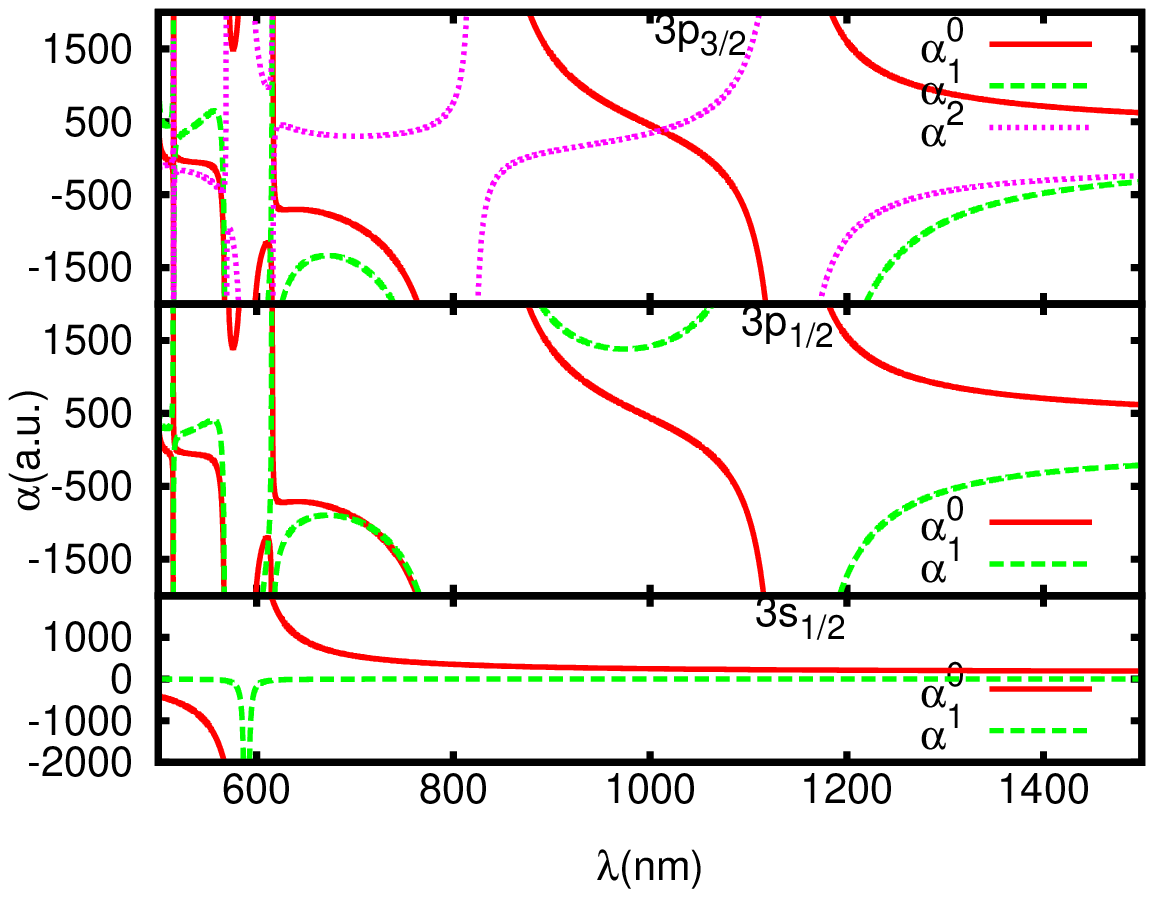}
  \caption{(color online) Scalar $\alpha^0$, vector $\alpha^1$ and tensor $\alpha^2$ polarizability (in a.u.) for $3s_{1/2}$, $3p_{1/2}$ and $3p_{3/2}$ state of Na atom as a function of wavelength (in n.m.)}
  \label{figna}     
\end{figure}

\begin{figure}[htp]
  \includegraphics[scale=0.6]{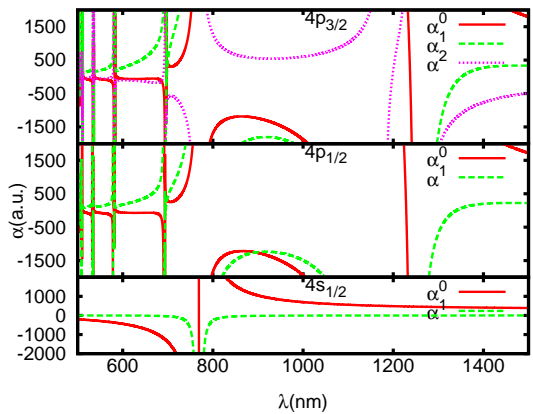}
  \caption{(color online) Scalar $\alpha^0$, vector $\alpha^1$ and tensor $\alpha^2$ polarizability (in a.u.) for $4s_{1/2}$, $4p_{1/2}$ and $4p_{3/2}$ state of K atom as a function of wavelength (in n.m.)}
  \label{figk}     
\end{figure}

\section{Results and Discussion}\label{sec4}
We first use the obtained E1 matrix elements and other MBPT(3) results to evaluate static
polarizabilities ($\omega=0$) of the considered states in Li, Na and K to bench mark their 
accuracies aganist the previously reported experimental and theoretical results. The matrix elements calculated using the method mentioned above are shown in Table~\ref{e1mat} and are presented under the column CCSD(T). Uncertainties
in the E1 matrix elements and MBPT(3) results are estimated using the procedures given in
\cite{sahoo-arora2, sahoo-nandy}. To reduce the uncertainties in our calculations, we have 
taken the E1 matrix elements complied in Ref.~\cite{volz}, they are pointed out in the table 
explicitly, instead from our calculations. Experimental energies from the national institute of
science and technology (NIST) database \cite{NIST,NIST1,NIST2} are used in the present calculations. The determined polarizabilities 
are given in Table \ref{pol} and compared with the other results. The 
most accurate experimental measurement of Li ground state polarizability 
$\alpha_{2s}^0 = 164.2(11)$ a.u. was obtained in \cite{li-exp}. Our result 
$\alpha_{2s}^0 = 164(1)$ a.u. is in excellent agreement with the experimental value. 
The most strigent experimental value for Na ground state polarizabilitty was obtained by 
interferometry experiment as $\alpha_{3s}^0 = 162.7(8)$ a.u.~\cite{na-exp} and our present 
value $\alpha_{3s}^0 = 162.4(2)$ a.u. agrees well with the experimental value within the 
uncertainty limits. The most recent experimental result available for the ground state 
polarizability in K is $\alpha_{4s}^0 = 290.58(1.42)$ a.u. ~\cite{k-exp}, which is very close to 
our calculated value $\alpha_{4s}^0 = 289.8(6)$ a.u. Our results for all the states in the 
considered three atoms also agree with other theoretical results. Therefore, the present
polarizability results can be used further to find out magic wavelengths in these atoms.

\begin{figure}[htp]
 \includegraphics[scale=0.6]{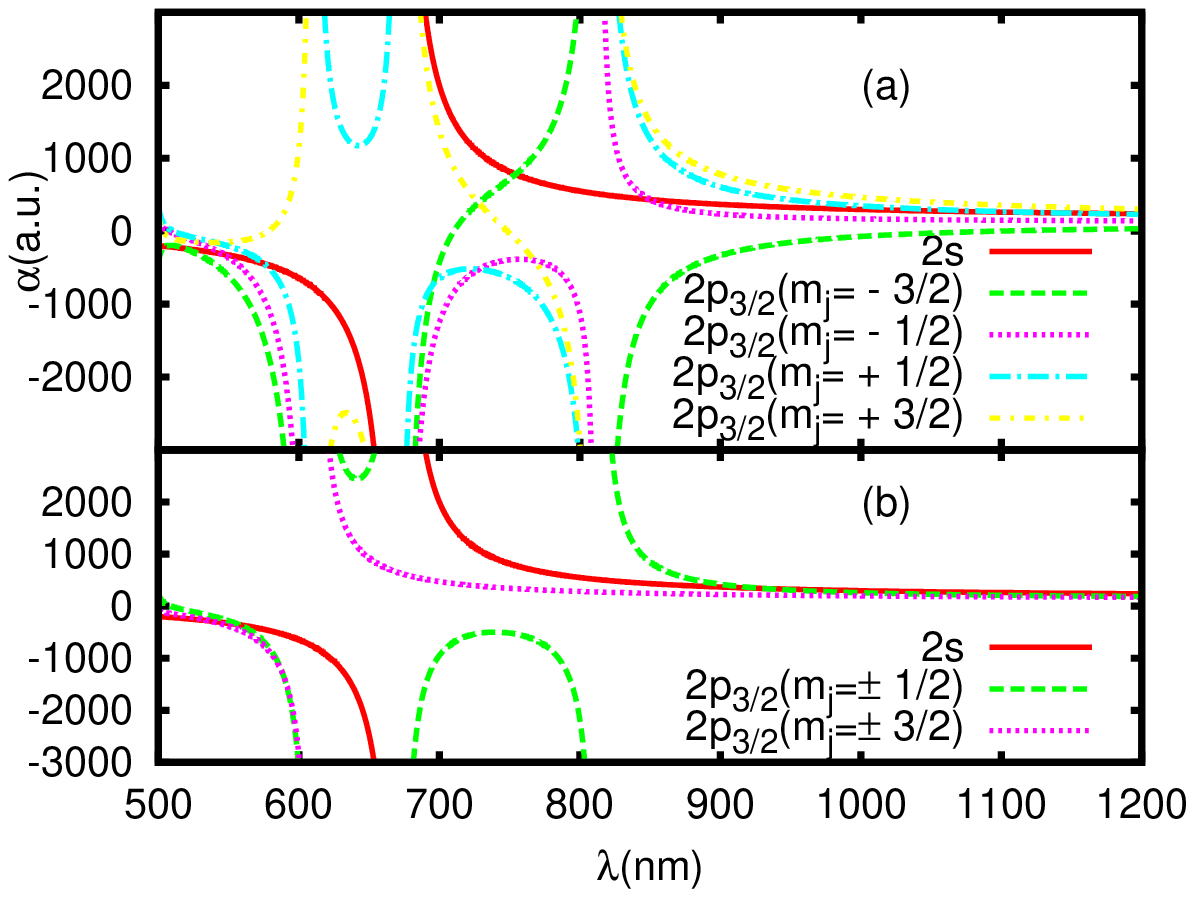}
  \caption{(color online)Total dynamic polarizability for $2s$ and $2p_{3/2}$ states of Li atom due to (a) linearly polarized light (b) circularly polarized light. Magic wavelength for the corresponding $2s-2p_{3/2}$ transition are given by the crossing of two polarizability curves.}
  \label{figli2}     
\end{figure}

\begin{figure}[htp]
  \includegraphics[scale=0.6]{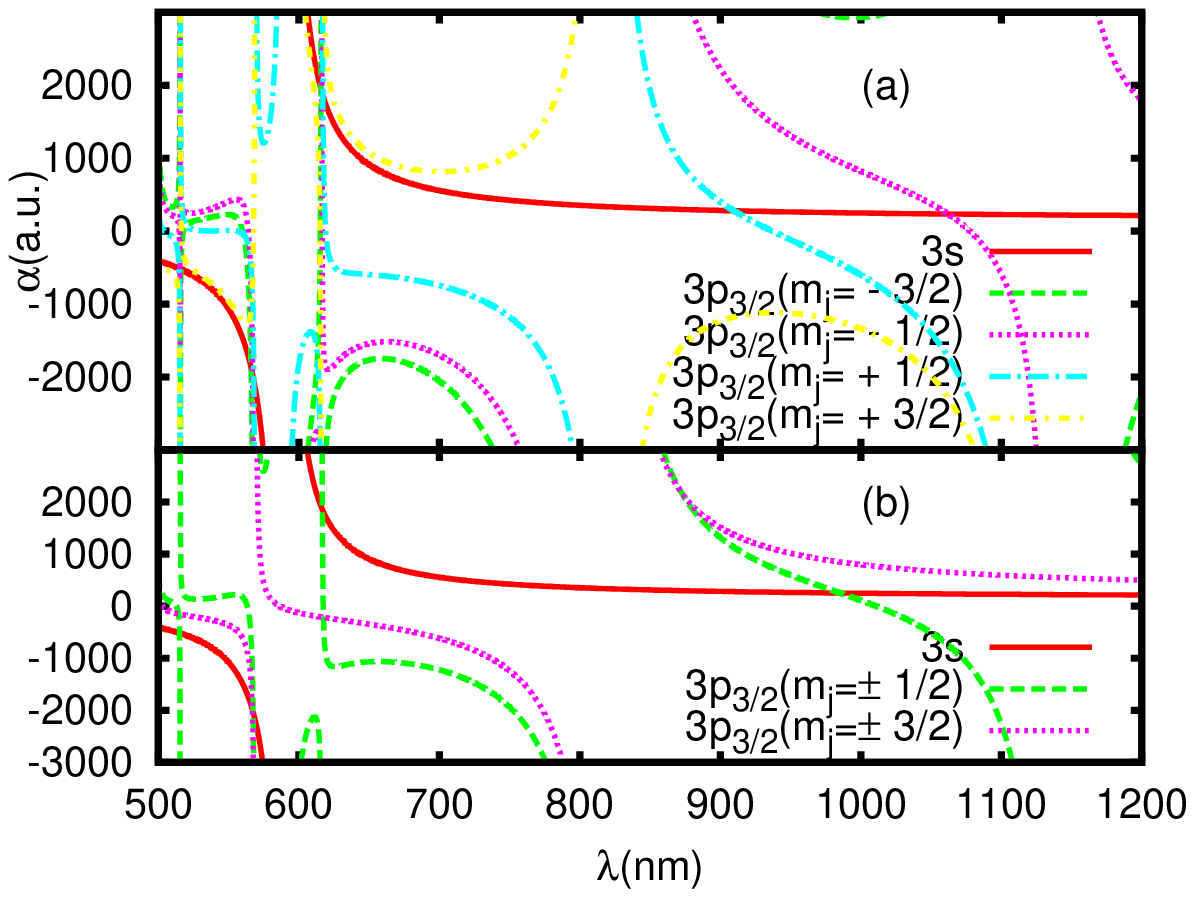}
  \caption{(color online) Total dynamic polarizability for $3s$ and $3p_{3/2}$ states of Na atom due to (a) linearly polarized light (b) circularly polarized light. Magic wavelength for the corresponding $3s-3p_{3/2}$ transition are given by the crossing of two polarizability curves.}
  \label{figna2}     
\end{figure}

\begin{figure}[htp]
  \includegraphics[scale=0.6]{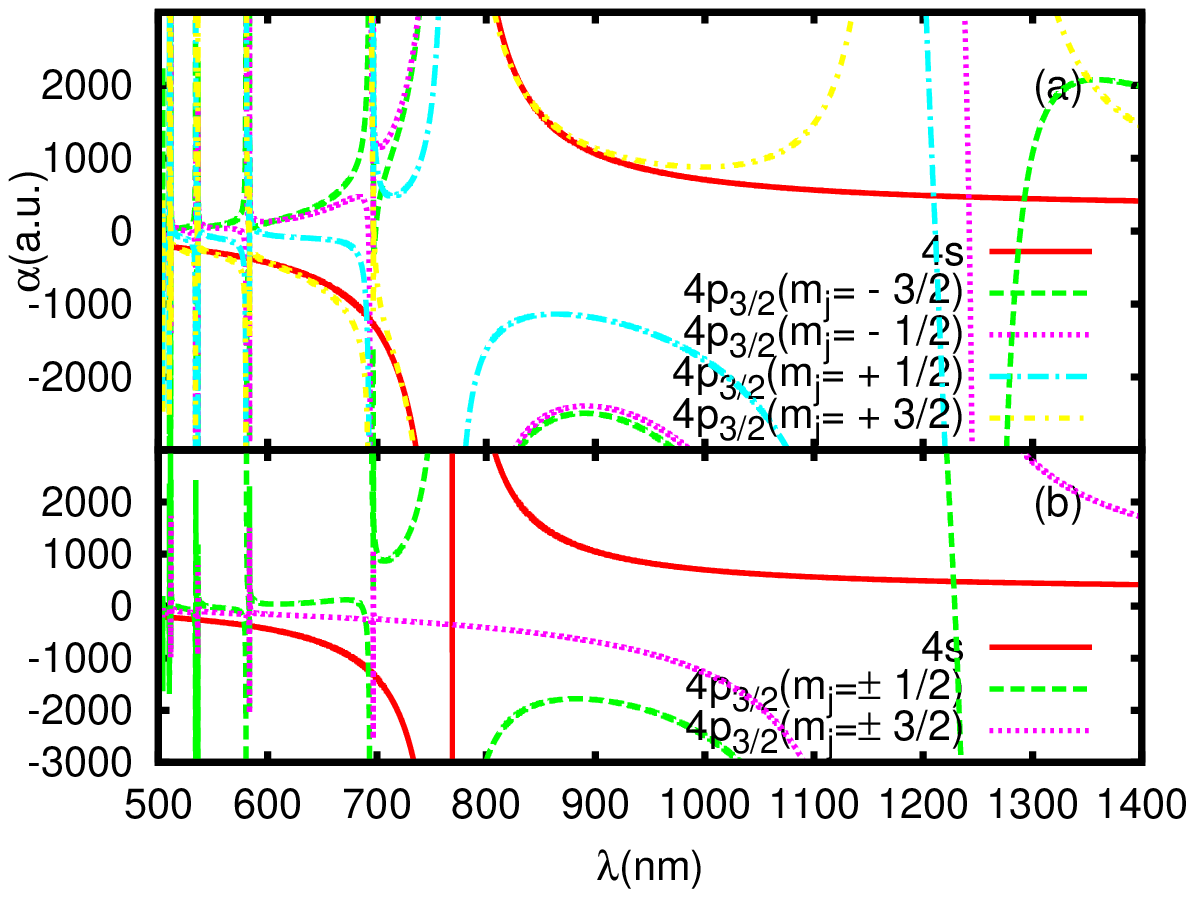}
  \caption{(color online) Total dynamic polarizability for $4s$ and $4p_{3/2}$ states of K atom due to (a) linearly polarized light (b) circularly polarized light. Magic wavelength for the corresponding $4s-4p_{3/2}$ transition are given by the crossing of two polarizability curves.}
  \label{figk2}     
\end{figure}

 To find magic wavelengths, i.e. the null differenial dynamic polarizabilities among various
states, we first determine the frequency dependent polarizabilities. In Figs.~\ref{figli}, 
\ref{figna}, and \ref{figk}, we present these results for the scalar, vector and tensor 
polarizabilities for Li, Na, and K atoms, respectively. In Figs.~\ref{figli2}, \ref{figna2}, 
and \ref{figk2} we plot the total dynamic polarizability for the $ns$ and $np_{3/2}$ states 
of Li, Na, and K atoms respectively (where $n=2$ for Li, $n=3$ for Na and $n=4$ for K). In the
case of the $np_{3/2}$ states, the total polarizability in the presence of linearly polarized light, 
is determined as $\alpha=\alpha_0-\alpha_2$ for  $m_j=\pm 1/2$ and $\alpha=\alpha_0+\alpha_2$  
for $m_j=\pm 3/2$. Similarly, the total polarizability for the $np_{3/2}$ states due to the circularly 
polarized light is determined separately for the $m_j=-3/2,-1/2,1/2,3/2$ sublevels using 
Eq.(\ref{cpl}). Magic wavelengths for the corresponding transitions are located at the crossing 
of the two curves.

\begin{table*}
\caption{\label{tabi} 
List of magic wavelengths in Li, Na and K atoms for the linearly polarized light.}
{\small
\begin{ruledtabular}
\begin{tabular}{lcccccc}
 & \multicolumn{2}{c}{$np_{1/2}\rightarrow ns$} & \multicolumn{4}{c}{$np_{3/2}\rightarrow ns$}\\ 
\multicolumn{1}{c}{$m_j\rightarrow$}   &    \multicolumn{2}{c}{$|1/2|$} & \multicolumn{2}{c}{$|1/2|$} & \multicolumn{2}{c}{$|3/2|$} \\
\hline
{\textbf{Li}}&Present&Ref.~\cite{safronova-li}& Present&Ref.~\cite{safronova-li}&Present&Ref.~\cite{safronova-li}\\
\hline
$\alpha(\lambda_{\rm{magic}})$ &-328 &-327.1(3)& -357& -357.26(7) & -288 & -288.0(3) \\
$\lambda_{\rm{magic}}$ & 550(1)& 549.42(6) & 557(1)& 557.16(2)& 537(3)& 537.61(7) \\

$\alpha(\lambda_{\rm{magic}})$ &-398&398.7(2)& 339& 339.9(2) & - & - \\
$\lambda_{\rm{magic}}$ & 873(2)& 872.57(9) & 931(2)& 930.3(2)& -&-  \\

$\alpha(\lambda_{\rm{magic}})$ &398.7(2) & 339.9(2) & -\\
$\lambda_{\rm{magic}}$ & 872.57(9) & 930.3(2)&-  \\
\hline
{\textbf{Na}}&Present&Ref.~\cite{arora1}& Present&Ref.~\cite{arora1}&Present&Ref.~\cite{arora1}\\
\hline
$\alpha(\lambda_{\rm{magic}})$ &-520&-514(1) &-522& -517(1) &-& - \\
$\lambda_{\rm{magic}}$ &514.73(1)& 514.72(1) & 515.01(1)& 515.01(1) & -&- \\

$\alpha(\lambda_{\rm{magic}})$ &-1981&-1956(3) &-2063& -2038(3) &-2002& -1976(3) \\
$\lambda_{\rm{magic}}$ &566.594(5)&566.57(1) & 567.451(3)& 567.43(1) & 566.819(2)& 566.79(1) \\

$\alpha(\lambda_{\rm{magic}})$ &53361 & 52760(100) &-66978& -66230(80) &-47 & -42(2) \\
$\lambda_{\rm{magic}}$ &589.4570(1)& 589.457 & 589.6363(1)& 589.636 & 589.5563(3) & 589.557(1)\\

$\alpha(\lambda_{\rm{magic}})$ &1931&1909(2) &1876& 1854(2)& - &- \\
$\lambda_{\rm{magic}}$ &615.872(1)& 615.88(1) & 616.708(1)& 616.712(1)&- & - \\

$\alpha(\lambda_{\rm{magic}})$ &244&241(1) &255& 252(1)& - &- \\
$\lambda_{\rm{magic}}$ &1028.6(4)& 1028.7(2)  & 984.7(9)& 984.8(1)& -  & - \\
\hline
{\textbf{K}}&Present&Ref.~\cite{arora1}& Present&Ref.~\cite{arora1}&Present&Ref.~\cite{arora1}\\
\hline
$\alpha(\lambda_{\rm{magic}})$ &-208&- &-209&- &-210&-\\
$\lambda_{\rm{magic}}$ & 502.59(1)&- & 503.70(2)&- & 504.0(1)&-\\

$\alpha(\lambda_{\rm{magic}})$ &-216& &-218& &-&\\
$\lambda_{\rm{magic}}$ & 508.086(4)&- &  509.31(5)&- &- &-\\

$\alpha(\lambda_{\rm{magic}})$ &-218 &-&-221&- &-221&-\\
$\lambda_{\rm{magic}}$ & 509.428(8)&- & 510.83(2)&- & 510.80(8)&-\\

$\alpha(\lambda_{\rm{magic}})$ &-258& &-260 &&-&\\
$\lambda_{\rm{magic}}$ & 532(2)&- & 533.07(1)& -& -&- \\

$\alpha(\lambda_{\rm{magic}})$ &-262&- &-266 &-&-265&-\\
$\lambda_{\rm{magic}}$ & 534.091(2)& -& 535.81(2)&-& 535.53(1)&-\\

$\alpha(\lambda_{\rm{magic}})$ &-367&- &-371 &-&-&-\\
$\lambda_{\rm{magic}}$ & 577.36(1)&- & 578.71(8)&- & -&-\\

$\alpha(\lambda_{\rm{magic}})$ &-379&- &-385&- &-384&-\\
$\lambda_{\rm{magic}}$ & 581.163(1)&- & 583.168(1)&- & 582.98(3)&-\\

$\alpha(\lambda_{\rm{magic}})$ &-1203&-1186(2)& - & - &- &-\\
$\lambda_{\rm{magic}}$ & 690.137(2)& 690.15(1) & - & -& - & - \\

$\alpha(\lambda_{\rm{magic}})$ &-1272&- &-1243& -1226(3) &-1331& -\\
$\lambda_{\rm{magic}}$ & 693.775(1)&  - & 692.31(2)& 692.32(2) & 696.582(1)& -\\

$\alpha(\lambda_{\rm{magic}})$ &21290& 20990(80) & -27566& -27190(60) &-356& -356(8) \\
$\lambda_{\rm{magic}}$ & 768.413(1)& 768.413(4) & 769.432(1)& 769.432(2) & 768.980(1)& 768.980(3)  \\

$\alpha(\lambda_{\rm{magic}})$ &479& 472(1) &479& 472(1)& -  &-\\
$\lambda_{\rm{magic}}$ & 1227.73(1)& 1227.7(2) & 1227.73(2)&1227.7(2) & -& -\\
\end{tabular}   
\end{ruledtabular}
}
\end{table*}

\begin{table*}
\caption{\label{tabc}
List of magic wavelengths for the circularly polarized light in Li, Na and K atoms.}
{\small
\begin{ruledtabular}
\begin{tabular}{lcccccc}
& \multicolumn{2}{c}{$np_{1/2}\rightarrow ns$} & \multicolumn{4}{c}{$np_{3/2}\rightarrow ns$}\\
$m_j\rightarrow$   &    $-1/2$ & $1/2$ & $-3/2$ & $-1/2$ & $1/2$ & $3/2$ \\
\hline\\
\textbf{Li} \\
$\alpha(\lambda_{\rm{magic}})$ &-274 & - & -239 & -316 & -449 & - \\
$\lambda_{\rm{magic}}$ & 533(1) & - & 518.8(2.5) & 546.1(1.3) & 575.5(0.9) & -\\

$\alpha(\lambda_{\rm{magic}})$ & 596 & - & 779 & 432 & 250 & - \\
$\lambda_{\rm{magic}}$ & 786.9(3)& - & 754.2(3) & 850.0(9) & 1140(5) & -\\

&& \\
\textbf{Na} \\
$\alpha(\lambda_{\rm{magic}})$ & -522 & -506 & -530 & -523 & -517 & - \\
$\lambda_{\rm{magic}}$ & 515.231(2)& 513.25(6) & 516.10(1)& 515.28(1) & 514.60(2) & -\\

$\alpha(\lambda_{\rm{magic}})$ & -1884 & - & -1893 & -1991 & -2076 & - \\
$\lambda_{\rm{magic}}$ & 565.864(5)& - & 565.963(7)& 567.076(4)& 567.947(3) & -\\

$\alpha(\lambda_{\rm{magic}})$ & 1994 & 1863 & 1974 & 1923 & 1878 & - \\
$\lambda_{\rm{magic}}$ & 615.412(1)& 617.352(7) & 615.694(1)& 616.435(1) & 617.117(2) & -\\

$\alpha(\lambda_{\rm{magic}})$ & 211 & - & - & 237 & 283 & - \\
$\lambda_{\rm{magic}}$ & 1252.9(8)& - &- & 1061.7(3) & 909.2(4) & -\\

&& \\
\textbf{K} \\
$\alpha(\lambda_{\rm{magic}})$ &- &-205 & - & -208 & -208 & -210 \\
$\lambda_{\rm{magic}}$ & - & 501.68(5) & - & 503.83(3) & 503.48(5) & 505.0(2)\\

$\alpha(\lambda_{\rm{magic}})$ &- &-212 & - & -218 & -216 & - \\
$\lambda_{\rm{magic}}$ & - & 506.17(6) & - & 509.73(1) & 508.71(2) & -\\

$\alpha(\lambda_{\rm{magic}})$ &-218 & -217 & -219.39 & -219.26 & -219.30 & -219.26 \\
$\lambda_{\rm{magic}}$ & 509.9(8) & 509.57(2) & -510.90(4) & 510.82(3) & 510.84(2) & 510.85(3))\\

$\alpha(\lambda_{\rm{magic}})$ &- &- & -219.8 & -219.9 & -220.0 & -220.5 \\
$\lambda_{\rm{magic}}$ & - & - & 511.168(1) & 511.20(8) & 511.298(4) & 511.56(3)\\

$\alpha(\lambda_{\rm{magic}})$ &-259 &-250 & - & -259 & -257 & -265 \\
$\lambda_{\rm{magic}}$ & 533.5(5) & 528.74(8) & - & 534(3) & 532.21(3) & 536.66(2)\\

$\alpha(\lambda_{\rm{magic}})$ &-374 &-348 & - & -370 & -363 & -382 \\
$\lambda_{\rm{magic}}$ &579(2) & 572.02(7) & - & 579.645(3) & 577.25(2) & 583.598(2)\\

$\alpha(\lambda_{\rm{magic}})$ &-1169 &-1170 & - & - & -& -\\
$\lambda_{\rm{magic}}$ &690.072(2) & 690.137(2) & - & - & - & -\\

$\alpha(\lambda_{\rm{magic}})$ &-1205 &-1069 & -1292 & -1228 & -1179 & -1294 \\
$\lambda_{\rm{magic}}$ &692.104(1) & 683.83(2) & 696.60976(1) & 693.329(1) & 690.63(6) & 696.69199(1)\\

$\alpha(\lambda_{\rm{magic}})$ &468 &- &  453 & 475& 491& -  \\
$\lambda_{\rm{magic}}$ & 1255.37(4) & -  & 1292.7(2) & 1241.6(1) & 1209.4(9) &-\\

\end{tabular}
\end{ruledtabular}
}
\end{table*}

\begin{table}
\caption{\label{tabavg1} 
List of average magic wavelengths $\lambda_{\rm{magic}}(\rm{avg})$ for Li, Na and K atoms for the linearly polarized light.}
\begin{ruledtabular}
\begin{tabular}{ccc}
&{$np_{1/2}\rightarrow ns$} & {$np_{3/2}\rightarrow ns$}\\ 
\hline
\bf{Li} \\
&{550(1)} & {547(20)} \\
&{873(2)} & {931(2)} \\
\\
\textbf{Na} \\
&{514.73(1)} & {515.01(1)} \\
&{566.59(1)} & {567(1)} \\
&{589.4570(1)} & {589.6(1)} \\
&{616.872(1)} & {616.708(1)} \\
&{1028.6(4)} & {984.7(9)} \\
\\
\textbf{K} \\
&{502.59(1)} & {503.8(3)} \\
&{508.086(4)} &{509.31(5)} \\
&{509.43(1)} & {10.8(1)} \\
&{532(2)} & {533.07(1)} \\
&{534.091(2)} & {535.7(3)} \\
&{577.36(1)} &{578.7(1)} \\
&{581.163(1)} &{583.1(8)} \\
&{690.137(2)} &{-} \\
&{693.7752(1)} &{694(4)} \\
&{768.413(1)} & {768.6(5)} \\
&{1227.73(1)} & {1227.73(2)} \\
\end{tabular}   
\end{ruledtabular}
\end{table}

\begin{table}
\caption{\label{tabavg2} 
List of average magic wavelengths $\lambda_{\rm{magic}}(\rm{avg})$ for Li, Na and K atoms for the left-circularly polarized light.}
\begin{ruledtabular}
\begin{tabular}{ccc}
&{$np_{1/2}\rightarrow ns$} &{$np_{3/2}\rightarrow ns$}\\
\hline\\
\textbf{Li} \\
& {533(1)} & {-} \\
& {786.9(3)} & {-} \\
 \\
\textbf{Na} \\
& {514(2)} &{515(2)} \\
& {565.86(1)} & {567(2)} \\
& {616(2)} & {616(1)} \\
& {1252.9(8)} &{985(153)} \\
\\
\textbf{K} \\
& {501.7(1)} & {504(1)} \\
& {506.2(1)} & {509(1)} \\
& {509.7(3)} & {510.9(1)} \\
& {-} & {511.3(4)} \\
& {531(5)} & {534(4)} \\
& {576(7)} & {580(6)} \\
& {690.1(1)} &{-} \\
& {688(4)} & {694(6)} \\
& {1255.37(4)} & {1248(84)} \\
\end{tabular}   
\end{ruledtabular}
\end{table}

In Table~\ref{tabi}, we list magic wavelengths for the considered atoms in the presence of linearly polarized
light. As shown in the table, the magic wavelengths found in the present work  agrees very well with the previous 
publications \cite{arora1,safronova-li}. 
We do not discuss these results in detail here since they are 
discussed in the above works and we focus mainly on the results obtained due to the circularly polarized light.
It is to be noted that we did not find any other data to compare our magic wavelength results in the case of K atom for wavelengths less than 600 nm. 
We consider hereafter the left-handed circularly polarized light for all the practical purposes as
the results will have a similar trend with the right-handed circularly polarized 
light due to the linear dependency of degree of polarizability $\mathcal{A}$ in Eq. (\ref{cpl}). 

In Table \ref{tabc}, we list magic wavelengths for the $ns-np_{1/2,3/2}$ transitions of Li, Na, and 
K atoms in the wavelength range 500$-$1500 $nm$ in the presence of circularly polarized light. As found, the number of magic wavelengths for the
$ns-np_{1/2}$ transitions for the circularly polarized light are less compared to the linearly
polarized light. Therefore, using the linearly polarized light to trap the atoms for these
transitions would be more advantage. However, the reported magic wavelengths could be useful in
a situation where it demands to trap the atoms using the circulalrly polarized light. Below, 
we discuss the results only for the $ns-np_{3/2}$ transitions as they seem to be of more experimental 
relevance owing to the fact that there are only fewer convenient magic wavelengths for these 
transitions found using the linearly polarized light. 

First we discuss the results for the $2s-2p_{3/2}$ transition of Li atom.
As seen in the table, the number of convenient magic wavelengths for the above transition in this atom is less compared to
the linearly polarized light. Moreover, no magic wavelength was located for the $m_j = 3/2$ sub-level. Therefore, it would
be appropriate to use linearly polarized light for the state insensitive trapping of this atom.  Next, we list a number of 
$\lambda_{\rm{magic}}$ and the corresponding polarizabilities for the $3s-3p_{3/2}$ transition of Na 
in the wavelength range 500$-$1500 $nm$ in the same table. The number inside the brackets for 
$\lambda_{\rm{magic}}$ depicts the uncertainty of the match of polarizability curves for the two 
states involved in the transition. These uncertainties are found as the maximum differences between 
the $\alpha_{3s} \pm \delta\alpha_{3s}$ and $\alpha_{3p} \pm \delta\alpha_{3p}$ contributions with 
their respective magnetic quantum numbers, where the $\delta\alpha$ are the uncertainties in the 
polarizabilities for their corresponding states. For Na atom, we get a set of four magic wavelengths 
in between six $3p_{3/2}$ resonances lying in the wavelength range 500$-$1400 $nm$; 
i.e. $3p_{3/2}-4s$ resonance at 1140.7 $nm$, $3p_{3/2}-3d_j$ resonance at 819.7 $nm$, $3p_{3/2}-5s$ resonance  at 616.3 $nm$, $3p_{3/2}-3s$ resonance at 589.2 $nm$, $5p_{3/2}-4d_j$ resonance at 569 $nm$, and $5p_{3/2}-6s$ resonance at 515.5 $nm$. 
The magic wavelength expected between $3p_{3/2}-5s$ and $3p_{3/2}-3s$ resonances is missing for the circularly polarized traps. 
Half of the magic wavelengths support blue-detuned whereas the other half favour towards the red-detuned optical traps. 
It can be observed from Table \ref{tabc} that $m_j=3/2$ sub-level does 
not support state-insensitive trapping at any of the listed magic wavelengths. 
However, using a switching trapping scheme 
as described in \cite{sahoo-arora2} can allow trapping this sub-level too.
The magic wavelength at 616 $nm$ is recommended owing to the
fact that  it supports a strong red-detuned trap 
as depicted by a large positive value of polarizability at this wavelength. 
For the $4s-4p_{3/2}$ transition in K atom, we get eight sets of magic wavelengths in the wavelength range $500-1500$ $nm$ as shown in Table~\ref{tabc}. 
Out of these eight magic wavelengths, the 
magic wavelength at 1247 $nm$ supports red-detuned optical trap.
 The magic wavelengths at 510.9, 511.3, and 694 $nm$ occur for all the $m_j$ sub-levels at nearly same value of polarizability. 
However, at  511.3 and 694 $nm$ the crossing for polarizability curves for the $4s$ and $4p_{3/2}$ states is very sharp. 
In addition to the magic wavelengths mentioned in Table \ref{tabc}, we found five more magic wavelengths 
for the $m_j=3/2$ state at 516.8(7), 543.4(5), 605.2(9), 724.5(3), and 849.7(8) $nm$. 

The final magic wavelengths are calculated as the average of the magic wavelengths for various $m_j$ sublevels and are written as $\lambda_{\rm{magic}}(\rm{avg})$ in Table \ref{tabavg1} and \ref{tabavg2}, for linearly and circularly polarized light respectively. The error in 
the  $\lambda_{\rm{magic}}(\rm{avg})$ is calculated as the maximum difference between 
the magic wavelengths from different $m_j$ sub-levels. 
For cases where the magic wavelength
was found for only one $m_j$ sublevel (for
example, $\lambda_{\rm{magic}}$ for the $4s-4p_{3/2}$ transition at 1227.73
nm for K atom) , the number in the bracket corresponds to
the uncertainty in the match of the polarizabilities of the $ns$
and $np$ states in place of representing the spread in the magic wavelengths for various 
$m_j$ sublevels.

\section{summary}
In summary, we have investigated 
magic wavelengths in Li, Na and K atoms for both the linearly and circularly polarized optical traps.
To determined these values, we have calculated dynamic polarizabilities using the best know E1
matrix elements. Our predictions for the linearly polarized trap agree well with the previously
reported results. This study demonstrates a significant number of magic wavelengths due to the
the circularly polarized light which will be very useful in trapping the 
above atoms in the ac Stark shift free regime. However, we do not recommend to use the circularly 
polarized traps for trapping Li atoms.

\section*{Acknowledgement} 
The work of B.A. was supported by the University Grants Commission and Department of Science and Technology, India. 
Computations were carried out using 3TFLOP HPC Cluster at Physical Research Laboratory, Ahmedabad.


\end{document}